\documentclass[a4paper,twoside]{article}

\usepackage{epsfig}
\usepackage{subcaption}
\usepackage{calc}
\usepackage{amssymb}
\usepackage{amstext}
\usepackage{amsmath}
\usepackage{amsthm}
\usepackage{multicol}
\usepackage{pslatex}
\usepackage{apalike}
\usepackage{algorithm2e}
\usepackage[bottom]{footmisc}

\usepackage{moreverb,url}
\usepackage{graphicx}
\usepackage{pifont}
\newcommand{\cmark}{\ding{51}}%
\newcommand{\xmark}{\ding{55}}%
\usepackage[english]{babel}
\usepackage{fancyhdr}
\usepackage{subfiles}
\usepackage{array}
\usepackage{multirow} 
\usepackage{mdframed}
\usepackage{booktabs}
\usepackage{color}
\usepackage{colortbl}
\definecolor{lightgray}{rgb}{0.9,0.9,0.9}

\usepackage{SCITEPRESS}     

\begin{document}

\title{Evaluating LLMs for Visualization Tasks}

\author{\authorname{Saadiq Rauf Khan, Vinit Chandak, Sougata Mukherjea}
\affiliation{Indian Institute of Technology, Delhi, India}
\email{saadiq351@gmail.com, defozex47@gmail.com, sougatam@iitd.ac.in}
}

\keywords{Large Language Models, Visualization Generation, Visualization Understanding}

\abstract{Information Visualization has been utilized to gain insights from complex data. In recent times, Large Language Models (LLMs) have performed very well in many tasks. In this paper, we showcase the capabilities of different popular LLMs to generate code for visualization based on simple prompts. We also analyze the power of LLMs to understand some common visualizations by answering simple questions. Our study shows that LLMs could generate code for some visualizations as well as answer questions about them. However, LLMs also have several limitations. We believe that our insights can be used to improve both LLMs and Information Visualization systems.}

\onecolumn \maketitle \normalsize \setcounter{footnote}{0} \vfill

\section{Introduction}
With the amount and complexity of information increasing at staggering rates, Information Visualization is being utilized to enable people understand and analyze information. Over the years many techniques have been developed for creating information visualizations of different types of data. Information visualization can be created using various tools\footnote{for example, Tableau: https://www.tableau.com}, libraries in many programming languages\footnote{for example, matplotlib: https://matplotlib.org/} as well as scripts\footnote{for example, VegaLite: https://vega.github.io/vega-lite/}. However, the complexity of these tools, libraries and scripts can pose a barrier, especially for individuals without a strong background in data science or programming. To address this, automation of visualization creation using artificial intelligence techniques has also been explored \cite{wu2022}.

Natural language interfaces allow users to generate visualizations using simple and intuitive commands. The integration of natural language processing in data visualization tools enhances the efficiency of data analysis. Analysts can now focus more on interpreting the data rather than the technicalities of creating visualizations. This advancement democratizes data analysis, making it more accessible to a broader audience.

Large Language Models (LLMs) like GPT-3 \cite{brown2020} are capable of completing text inputs to produce human-like results. They have revolutionized Natural Language Processing by achieving state-of-the-art results on various tasks. Similarly, deep learning models that are trained on a large amount of existing code and can generate new code given some forms of specifications such as natural language descriptions or incomplete code \cite{chen2021}. 

Another important task is the machine understanding of the visualizations. It accelerates data analysis by allowing machines to process and interpret large volumes of visual data quickly, reducing the time needed for manual interpretation. Moreover, it improves accuracy by providing consistent extraction of information from visualizations. 

In this paper, we explore whether visualizations can be created or understood by prompting Large Language Models in natural language. Given the enormous potential of LLMs our aim was to explore whether LLMs are ready for Visualization tasks. Firstly, we evaluated whether popular LLMs like OpenAI's GPT-4\footnote{gpt4: https://https://openai.com/index/gpt-4}, Google's Gemini\footnote{Gemini: https://gemini.google.com} and Anthropic's Claude\footnote{Claude: https://www.anthropic.com/claude} could generate code for visualizations based on some simple prompts. Secondly, we investigated whether the LLMs could understand simple visualizations and answer questions about them. Our analysis shows that for some tasks LLMs performed very well; for example, most LLMs could produce code to generate simple visualizations. However, our study has also exposed several limitations of the LLMs - they were incorrect in several tasks - both in generation and understanding. 

The two main contributions of the paper are as follows:
\begin{enumerate}
    \item We have done an analysis of the capabilities of some of the popular LLMs to generate Python code and Vega-lite scripts for visualizations based on prompts.
    \item We explored the power of LLMs to understand simple visualizations and answer questions about them.
\end{enumerate}

The remainder of the paper is organized as follows. Section 2 cites related work. Section 3 analyzes the LLMs for visualization generation, while Section 4 analyzes the LLMs for Visualization Understanding. Finally, Section 5 concludes the paper.
\section{Related Work}
\subsection{Large Language Models}
Large Language Models like GPT-3 \cite{brown2020} have shown impressive results in various natural language understanding tasks. Given a suite of appropriate prompts\footnote{A system prompt is a set of fixed instructions created by the developers to constrain the LLM's response} a single LLM can be used to solve a great number of tasks. Various prompt engineering techniques have been developed to find the most appropriate prompts to allow a LLM to solve the task at hand \cite{liu2023}. On the other hand, Codex which is trained on 54 million software repositories on GitHub, has demonstrated stunning code generation capability — solving over 70\% of 164 Python programming tasks with 100 samples \cite{chen2021}.

\subsection{Visualization Generation}
With the popularity of information visualization, many techniques have been developed to create visualizations for different types of data. Information visualization can be created using various tools, libraries in many languages, as well as scripts based on Visualization Grammars. 

AI techniques have also been explored to automate the creation of visualizations, for example, using decision trees \cite{wang2020} and sequence-to-sequence recurrent neural networks \cite{dibia2019}. ChartSpark \cite{xiao2024} is a pictorial visualization authoring tool conditioned on both semantic contexts conveyed in textual inputs and data information embedded in plain charts. 

One significant direction of research is automating the creation of data visualizations based on users' natural language queries. Many systems for using natural language to generate visualizations (NL2VIS) are based on libraries of natural language processing. For example, NL4DV \cite{nar2021} uses CoreNLP \cite{man2014}. These systems either have constraints on user input or cannot understand complex natural language queries \cite{she2023}. Researchers have also trained neural networks using deep learning-based approaches \cite{luo2022} to process complex natural languages. However, a single approach based on deep learning cannot perform well on various tasks.

With the popularity of LLMs, there is significant interest in their application across various fields, including data visualization. \cite{vaz2024} investigates the capabilities of ChatGPT in generating visualizations. This study systematically evaluates whether LLMs can correctly generate a wide variety of charts, effectively use different visualization libraries, and configure individual charts to specific requirements.  The study concludes that while ChatGPT show promising capabilities in generating visualizations, there are still areas needing improvement. 

Similarly, \cite{li2024} explores the capabilities of GPT-3.5, to generate visualizations in Vega-Lite from natural language descriptions using various prompting strategies. The key findings reveal that GPT-3.5 significantly outperforms previous state-of-the-art methods in the NL2VIS task. It demonstrates high accuracy in generating correct visualizations for simpler and more common chart types. However, the model struggles with more complex visualizations and tasks that require a deeper understanding of the data structures. 

 LLMs have been integrated into NL2VIS systems, such as Chat2Vis \cite{mad2023} and LIDA \cite{dib2023}, which generate Python code to construct data visualizations. However, there remains a need for a systematic evaluation of how well these LLMs can generate visualizations using different prompt strategies.

\subsection{Visualization Understanding}
In recent times various Multi-modal Large Language models (MMLLMs) have been proposed for understanding of charts. Examples include ChartAssistant \cite{men2024} and UReader \cite{ye2023}. Many datasets and benchmarks have also been introduced  to test the capabilities of LLMs and MMMLLMs for chart understanding. Examples include ChartQA \cite{mas2022} and HallusionBench \cite{gua2024}. 
Research has also been done to evaluate the Large Language models in different aspects of visualization understanding. For example, \cite{ben2025} evaluates GPT-4 for various visualization literacy tasks, including answering questions and identifying deceptive visualizations. The assessment finds that GPT-4 can perform some tasks very efficiently, but struggles with some other tasks.

\section{Analyzing LLMs for Visualization Generation}
\begin{table*}[h!]
\centering
 \begin{tabular}{| p{4cm} | l | l | l | l |} 
 \hline
 Chart Type & GPT-3.5 & GPT-4o & Gemini & Claude \\ [0.5ex] 
 \hline\hline
Area Chart & \cmark & \cmark & \cmark & \cmark \\
Bar Chart & \cmark & \cmark & \cmark & \cmark \\
Box Plot & \cmark & \cmark & \cmark & \cmark \\
Bubble Chart & \cmark & \cmark & \cmark & \cmark \\
Bullet Chart & \xmark & \cmark & \xmark & \xmark \\
Choropleth & \cmark & \cmark & \cmark & \xmark \\
Column Chart & \cmark & \cmark & \cmark & \cmark \\
Donut Chart & \cmark & \cmark & \cmark & \cmark \\
Dot Plot & \xmark & \cmark & \cmark & \cmark \\
Graduated Symbol Map & \xmark & \cmark & \xmark & \xmark \\
Grouped Bar Chart & \cmark & \cmark & \cmark & \cmark \\
Grouped Column Chart & \cmark & \cmark & \cmark & \cmark \\
Line Chart & \cmark & \cmark & \cmark & \cmark \\
Locator Map & \cmark & \cmark & \cmark & \cmark \\
Pictogram Chart & \xmark & \xmark & \xmark & \xmark \\
Pie Chart & \cmark & \cmark & \cmark & \cmark \\
Pyramid Chart & \xmark & \cmark & \xmark & \xmark \\
Radar Chart & \cmark & \cmark & \xmark & \xmark \\
Range Plot & \cmark & \cmark & \xmark & \xmark \\
Scatter Plot & \cmark & \cmark & \cmark & \cmark \\
Stacked Bar Chart & \cmark & \cmark & \cmark & \cmark \\
Stacked Column Chart & \cmark & \cmark & \cmark & \cmark \\
Violin Plot & \cmark & \cmark & \cmark & \cmark \\
XY Heatmap Chart & \cmark & \cmark & \cmark & \cmark \\
\hline
Total & 19(79\%) & 23(95\%) & 18(75\%) & 17(70\%)\\ [0.5ex] 
 \hline
 \end{tabular}
 \caption{Performance Comparison of LLMs in Chart Generation using Python}
\label{table:default}
\end{table*}

\begin{figure*}
    \begin{center}
    \includegraphics[height=4in]{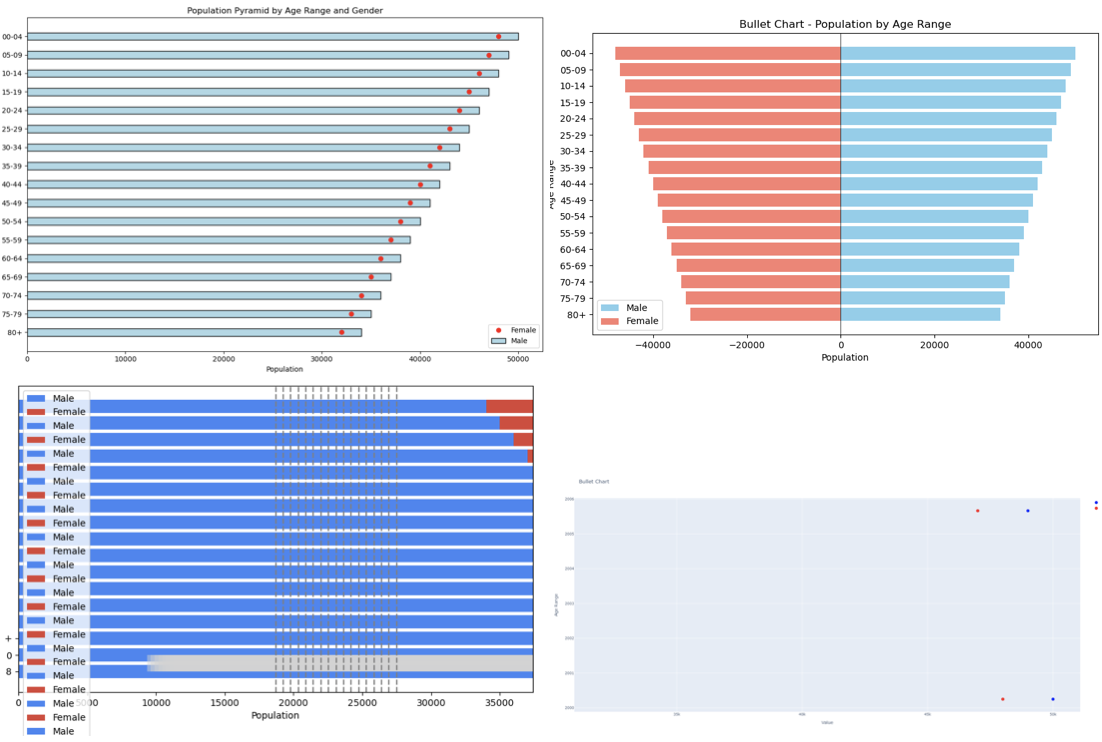}
    \caption{Comparison of Bullet charts. Only GPT-4o (top left) was able to produce the correct chart. GPT 3.5 produced a pyramid chart instead (top right). Gemini's and Claude's outputs were erroneous (bottom).}
    \label{fig:bullet}
    \end{center}
\end{figure*}

\begin{table}[h!]
\centering
 \begin{tabular}{| l | l | l | } 
 \hline
 Chart Type & GPT-4o & Gemini \\
 \hline
Area Chart & \cmark & \cmark \\
Bar Chart & \cmark & \cmark  \\
Box Plot & \cmark & \cmark  \\
Bubble Chart & \cmark & \xmark \\
Bullet Chart & \xmark & \xmark \\
Choropleth & \cmark & \xmark \\
Column Chart & \cmark & \cmark \\
Donut Chart & \cmark & \cmark \\
Dot Plot & \cmark & \xmark \\
Graduated Symbol Map & \cmark & \xmark \\
Grouped Bar Chart & \xmark & \xmark \\
Grouped Column Chart & \xmark & \xmark \\
Line Chart & \cmark & \xmark \\
Locator Map & \cmark & \xmark \\
Pictogram Chart & \xmark & \xmark \\
Pie Chart & \cmark & \cmark \\
Pyramid Chart & \cmark & \xmark  \\
Radar Chart & \xmark & \xmark \\
Range Plot & \xmark & \xmark \\
Scatter Plot & \cmark & \cmark \\
Stacked Bar Chart & \cmark & \cmark\\
Stacked Column Chart & \cmark & \cmark \\
Violin Plot & \xmark & \xmark \\
XY Heatmap Chart & \cmark & \cmark \\
\hline
Total & 17(70\%) & 10(41\%) \\ 
 \hline
 \end{tabular}
 \caption{Performance Comparison of LLMs in Chart Generation using Vega-lite scripts}
\label{table:vega}
\end{table}
\subsection{Process}
To evaluate the capabilities of LLMs in generating information visualizations, we followed a similar process as \cite{vaz2024}. We prompt the LLM to create a visualization based on a given specification and examine the code generated by the LLM. We chose Python for generating the visualization code due to its wide array of visualization libraries like \textit{matplotlib}. We also examine the ability of the LLMs to generate \textit{Vega-lite} scripts.

The methodology for the analysis involved several key steps:
\begin{enumerate}
\item
Selection of Visualization Techniques: We selected 24  visualization techniques for tabular data. These include common charts like bar graphs and pie charts as well as charts that may not be that popular like Violin Plots and Locator Maps. We exclude visualization techniques for hierarchical and network representations.
\item Creation or Acquisition of Suitable Datasets: We created or sourced data sets that were appropriate for the chosen visualization techniques, ensuring that they provided a robust basis for testing. These data sets cover a wide range of data types, including categorical, quantitative, temporal, and geographical data. This enables a comprehensive evaluation of the LLMs' ability to generate accurate and varied visualizations. 
\item 
Selection of LLMs to Analyze: We utilized 4 LLMs - OpenAI's GPT-3.5 and GPT-4o  as well as Google's Gemini-1.5-pro and Anthropic's Claude 3 Opus for our analysis to provide a broad perspective on the capabilities of current popular models generalizable across different LLM designs.
\item 
Design and Fine-tuning of Prompts: We used zero-shot prompting\footnote{Zero-shot prompting is a machine learning technique that involves giving an AI model a task or question without providing any specific training or examples \cite{liu2023}} for this task. We carefully designed and refined the prompts to maximize the effectiveness and accuracy of the LLMs in generating the desired visualizations. An example prompt is:
\textit{Can you write a Python script that generates a Bubble chart using columns mpg (quantitative), disp (quantitative), and hp (quantitative) from the CSV file cars.csv?}
\item 
Testing: We conducted a thorough test to evaluate the performance of LLMs, examining the variety of charts they could generate.
\end{enumerate}

\subsubsection{Experimental Procedure}
The assessment of the LLMs focused on the accuracy, efficiency, and versatility of the models in producing effective visual representations of data. For each experiment, we followed the following process to ensure consistency and accuracy:
\begin{enumerate}
\item
Initialize a New Session: Begin each experiment by creating a fresh session. Given that LLM chat sessions utilize previous prompts as context, it was crucial to start with a new session for each experiment. This approach ensured that each test was conducted independently, preventing any carry-over effects from previous prompts. For example, if multiple prompts requested charts in Vega-lite, subsequent prompts without a specified library or language might default to Vega-lite.
\item
Consistent Prompt Input: Enter all prompts within the same session and on the same day to maintain uniform conditions. 
\item  Execute and Analyze: Utilize the LLM output (either Python code or Vega-lite scripts) to create a visualization and analyze it. 
\end{enumerate}

\subsection{Chart generation using Python}
In the first analysis, each of the 4 LLMs was prompted to generate Python code for all the 24 distinct chart types. The performance of the LLMs are shown in Table \ref{table:default}. Each tick mark (\checkmark) represents a correct generation and each cross mark (\xmark) represents an incorrect generation. GPT-4o came out to be the best performer with the ability to produce around 95\% of the charts followed by GPT-3.5 being able to produce 79\% of the charts. The performance of Gemini and Claude was similar to that of GPT 3.5. 

Note that correct generation means that the LLM could produce correct code for the visualization based on the requirement specified by the prompt. Since LLMs are known to produce inconsistent results, we tuned the LLM parameters so that randomness in the output is minimized. During the experiments each prompt is repeated three times and we accept the outputs only if they remain the same.

Most of the errors were due to the lack of knowledge of some LLMs on certain types of visualization, especially uncommon ones. For example, only GPT-4o was able to produce the correct bullet charts. GPT 3.5 produced a Pyramid chart instead, whereas Gemini's and Claude's outputs were erroneous. The comparison is shown in Figure \ref{fig:bullet}. 

\subsection{Chart generation via Vega-lite scripts}
We also wanted to test and compare the performance of the aforementioned LLMs to generate Vega-lite scripts. Here we prompted the LLMs to generate Vega-lite scripts for all the 24 selected charts. For this evaluation, we used GPT-4o and Gemini for experimentation.

\begin{figure*}
    \begin{center}
    \includegraphics[height=2.5in]{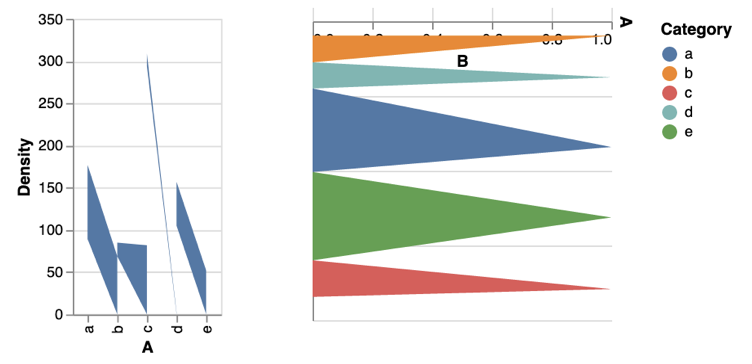}
    \caption{Both GPT-4o and Gemini could not produce Violin charts via Vega-lite scripts.}
     \label{fig:violin-comp}
    \end{center}
\end{figure*}

For the Vega-lite scripts, the results are shown in Table \ref{table:vega}. Vega-Lite proved to be difficult for LLMs. Gemini was  only able to create roughly 40\% of the total charts. The performance of GPT-4o also reduced significantly when switching from Python to Vega-lite. For example, both GPT-4o and Gemini could not produce Violin charts as shown in Figure \ref{fig:violin-comp}.

\section{Analyzing LLMs for Visualization Understanding}
\subsection{Data Set}
For analyzing the capabilities of LLMs for understanding visualizations, we have used the FigureQA dataset \cite{kah2018}. FigureQA consists of common charts accompanied by questions and answers concerning them. The corpus is synthetically generated on a large scale: its training set contains $100,000$ images with $1.3$ million questions. The corpus has five common visualizations for tabular data, namely, horizontal and vertical bar graphs, continuous and discontinuous line charts, and pie charts. 

 There are 15 types of questions that compare the quantitative attributes of two plot elements or one plot element with all the others. In particular, the questions examine properties such as the maximum, minimum, median, roughness, and greater than/less than relationships. All are posed as a binary choice between yes and no.

\begin{table*}[h!]
\centering
\begin{tabular}{>{\bfseries}l r r r}
\toprule
Metric                              & Gemini-1.5-pro  & GPT-4o   & Claude 3 Opus       \\
\midrule
Total Questions                     & 1,342           & 1,342     &1,342        \\
Total Correct Answers               & 863             & 886       &733        \\
Total Wrong Answers                 & 479             & 456       &609        \\
Accuracy (\%)                       & 64.31\%         & 66.02\%   &54.61\%        \\
\bottomrule
\end{tabular}
\caption{Comparison of Performance Metrics between LLMs to answer Yes-No (binary) questions}
\label{tab:model_comparison}
\end{table*}
\subsection{Automated Analysis on FigureQA}
To evaluate the ability of LLMs to understand and answer questions of information visualization we randomly chose $100$ images from the data set and the corresponding $1,342$ questions. Our random choice of the images will lead to variations in the chart types. We evaluated 3 LLMs -  Google's Gemini-1.5-pro, OpenAI's GPT-4o and Anthropic's Claude 3 Opus. The results of the evaluation are shown in Table \ref{tab:model_comparison}. As we can see, GPT-4o is the best performer, followed by Gemini-1.5-pro which is slightly behind and then Claude 3 Opus, which is much worse when compared to the other two models. 

\subsection{Need for Manual Analysis}
While this initial automated test with the FigureQA dataset provided quantitative metrics for evaluating the performance of the selected LLMs, we know that relying solely on binary questions does not offer a comprehensive assessment of the model's true comprehension abilities. The binary nature of the FigureQA questions introduces a significant limitation: the susceptibility to random guessing. Models can achieve approximately $50\%$ accuracy by making random choices without genuinely understanding the content of the figure/chart. 

\subsection{Data for Manual Analysis}
To address this limitation, we moved beyond automated binary questioning and incorporated manual analysis as a crucial step in our methodology. This involved developing custom, non-binary questions aimed at probing deeper into the visual reasoning abilities of the models. For the manual analysis, we have selected $20$ random charts for each of chart type in the FigureQA dataset. We have introduced new non-binary questions for each chart type that are useful to evaluate the level of understanding a model has of a chart. Examples of these questions are:
\begin{itemize}
\item {\bf Vertical/Horizontal Bar Chart:}  How many bars are there? 
\item {\bf Line Chart:} How many dotted/non-dotted lines are there? 
\item {\bf Pie Chart:}  Which color pie has the largest area?
\end{itemize}

\begin{table*}[ht]
\centering
\renewcommand{\arraystretch}{1.5} 
\setlength{\arrayrulewidth}{0.8pt} 
\setlength{\extrarowheight}{2pt} 
\begin{tabular}{|>{\centering\arraybackslash}m{6cm}|c|c|c|}
\hline 
\textbf{} & \textbf{Gemini 1.5 Pro} & \textbf{GPT-4o} & \textbf{Claude 3 Opus} \\ 
\hline
Images for which all qs answered correctly without prompt. & $53.8\%$ & $51.3\%$ & $33.8\%$ \\ 
\hline
Images for which all qs answered correctly with prompt. & $63.8\%$ & $57.5\%$ & $38.8\%$ \\ 
\hline
Qs answered without prompt. & $84.8\%$ & $87\%.8$ & $70.5\%$ \\ 
\hline
Qs answered correctly with prompt. & $89.2\%$ & $89.4\%$ & $73.4\%$ \\ 
\hline
\end{tabular}
\caption{Comparison of Performance Metrics between LLMs to answer non Yes-No (non-binary) questions}
\label{table:manual}
\end{table*}

\begin{figure}[h]
      \begin{center}
          \includegraphics[height=2in]{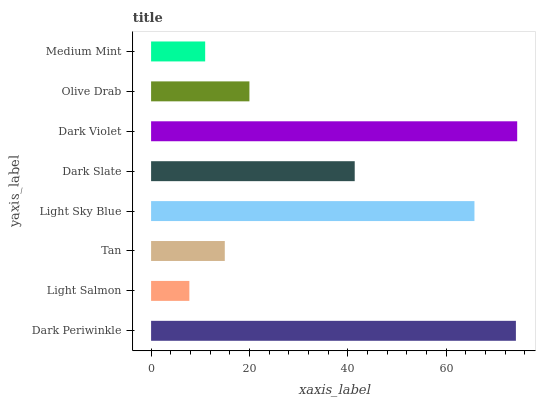} 
        \caption {All three models could not determine the color of the longest bar since all of the models struggle in determining the larger/smaller bars}
      \label{fig:close_hbar}
      \end{center}
\end{figure}

\subsection{Manual Analysis Results}
For each LLM, for each of the chart, we uploaded the image and then asked it all the questions. All the answers were checked against the correct  answers for the same questions. This, along with the inter-LLM comparison, will inherently compare the LLMs performance with a human baseline. We have also compared the models performance with and without a simple system prompt:
\textit{Analyze the following chart carefully and answer the following questions correctly.}

Table \ref{table:manual} shows the results. GPT-4o and Gemini performance were almost identical and better than Claude's performance. From the analysis we gained several key insights as follows:
\begin{itemize}
    \item \textbf{Performance Across Different Chart Types}
    \begin{itemize}
        \item The performance of LLMs varied significantly among different types of charts. For example, GPT-4o had $85$\% accuracy on pie charts and a mere $20$\% accuracy on line charts. This suggests that certain visualization formats may be easier for machines to interpret than others.
        \item Most of the models performed much better on pie charts as compared to other chart types. 
        \item All of the models performed very poorly on the line charts. This might be because of the presence of dotted lines, which might be treated as some kind of noise by the models. Sometimes Gemini-1.5-Pro did not recognize the dotted lines at all - especially when there is a mixture of dotted and non-dotted lines,
        \item All the models struggled with identifying relationships between close boundaries and lengths of shapes. When the bar lengths are close on a bar graph, all of the models struggled in comparing them; An example is shown in Figure \ref{fig:close_hbar}).
    \end{itemize}
    \item \textbf{Impact of System Prompts}
    \begin{itemize}
        \item In all the cases, the use of system prompts improved the performance of models. The improvement varied with the models and the chart types. Gemini-1.5-Pro improved significantly with the use of system prompts.
        \item This shows the importance of context and guidance in improving the performance of LLMs.
    \end{itemize}
    \end{itemize}

\section{Conclusion}
In this paper, we explore the capabilities of LLMs in generating visualizations from natural language commands. We evaluated the performance of various prominent LLMs in creating different types of chart using Python and Vega-Lite scripts. In addition, we analyze the abilities of LLMs in understanding and answering questions about some charts. The paper extends the prior art to explore the capabilities of LLMs for visualization generation and understanding. The findings of our research provide valuable insight into the current state of LLMs in the field of data visualization. Our study shows that LLMs are very efficient in some tasks, but fail in some more complex tasks. The results of this paper can be used to address the limitations of LLMs and improve them in the future. 
Some areas of future work include the following.
\begin{itemize}
\item We want to explore whether more advanced prompting techniques like Chain-of-Thought \cite{wei2022} can improve the results. 
\item We need to expand the analysis to other types of information visualization such as graphs and trees.
\item Combining the capabilities of LLMs and visualization tools to generate interactive visualizations is another promising research direction.
\end{itemize}

\bibliographystyle{apalike}
{\small
\bibliography{refs}}

\begin{thebibliography}{}

\bibitem[Bendeck and Stasko, 2025]{ben2025}
Bendeck, A. and Stasko, J. (2025).
\newblock {An Empirical Evaluation of the GPT-4 Multimodal Language Model on Visualization Literacy Tasks}.
\newblock {\em IEEE Transactions on Visualization and Computer Graphics}, 31(1):1105--1115.

\bibitem[Brown et~al., 2020]{brown2020}
Brown, T., Mann, B., Ryder, N., Subbiah, M., Kaplan, J.~D., et~al. (2020).
\newblock {Language Models are Few-shot Learners}.
\newblock In {\em Proceedings of the 34th International Conference on Neural Information Processing Systems (NIPS '20)}, pages 1877--1901.

\bibitem[Chen et~al., 2021]{chen2021}
Chen, M., Tworek, J., Jun, H., Yuan, Q., Ponde, H., et~al. (2021).
\newblock {Evaluating Large Language Models Trained on Code}.
\newblock {\em ArXiv}.

\bibitem[Dibia, 2023]{dib2023}
Dibia, V. (2023).
\newblock {LIDA: A Tool for Automatic Generation of Grammar agnostic Visualizations and Infographics using Large Language Models}.
\newblock In {\em Proceedings of the 61st Annual Meeting of the Association for Computational Linguistics (Volume 3: System Demonstrations), 2023}, pages 113 -- 126.

\bibitem[Dibia and Demiralp, 2019]{dibia2019}
Dibia, V. and Demiralp, C. (2019).
\newblock {Data2Vis: Automatic Generation of Data Visualizations Using Sequence-to-Sequence Recurrent Neural Networks}.
\newblock {\em IEEE Computer Graphics and Applications}, 39(5):33--46.

\bibitem[Guan et~al., 2024]{gua2024}
Guan, T., Liu, F., Wu, X., Xian, R., Li, Z., et~al. (2024).
\newblock {HallusionBench: An Advanced Diagnostic Suite for Entangled Language Hallucination and Visual Illusion in Large Vision-Language Models}.
\newblock In {\em Computer Vision and Pattern Recognition (CVPR 2024)}.

\bibitem[Kahou et~al., 2018]{kah2018}
Kahou, S.~E., Michalski, V., Atkinson, A., Kadar, A., Trischler, A., and Bengio, Y. (2018).
\newblock {FigureQA: An Annotated Figure Dataset for Visual Reasoning}.
\newblock {\em ArXiv}.

\bibitem[Li et~al., 2024]{li2024}
Li, G., Wang, X., Aodeng, G., Zheng, S., Zhang, Y., Ou, C., Wang, S., and Liu, C.~H. (2024).
\newblock {Visualization Generation with Large Language Models: An Evaluation}.
\newblock {\em ArXiv}.

\bibitem[Liu et~al., 2023]{liu2023}
Liu, P., Yuan, W., Fu, J., Jiang, Z., Hayashi, H., and Neubig, G. (2023).
\newblock {Pre-train, Prompt, and Predict: A Systematic Survey of Prompting Methods in Natural Language Processing}.
\newblock {\em ACM Computing Surveys}, 55(9):1--35.

\bibitem[Luo et~al., 2022]{luo2022}
Luo, Y., Tang, N., Li, G., Tang, J., Chai, C., and Qin, X. (2022).
\newblock {Natural Language to Visualization by Neural Machine Translation}.
\newblock {\em IEEE Transactions on Visualization and Computer Graphics}, 28(1):217--226.

\bibitem[Maddigan and Susnjak, 2023]{mad2023}
Maddigan, P. and Susnjak, T. (2023).
\newblock {Chat2Vis: Generating Data visualizations via Natural Language using Chatgpt, Codex and GPT-3 Large Language Models}.
\newblock {\em IEEE Access}, 11.

\bibitem[Manning et~al., 2014]{man2014}
Manning, C.~D., Surdeanu, M., Bauer, J., Finkel, J.~R., Bethard, S., and McClosky, D. (2014).
\newblock {The Stanford CoreNLP Natural Language Processing Toolkit}.
\newblock In {\em Proceedings of 52nd Annual Meeting of the Association for Computational Linguistics (2014): System Demonstrations}, pages 55--60.

\bibitem[Masry et~al., 2022]{mas2022}
Masry, A., Do, X.~L., Tan, J.~Q., Joty, S., and Hoque, E. (2022).
\newblock {{ChartQA: A Benchmark for Question Answering about Charts with Visual and Logical Reasoning}}.
\newblock In {\em Findings of the Association for Computational Linguistics: ACL 2022}, Dublin, Ireland.

\bibitem[Meng et~al., 2024]{men2024}
Meng, F., Shao, W., Lu, Q., Gao, P., Zhang, K., Qiao, Y., and Luo, P. (2024).
\newblock {ChartAssisstant: A Universal Chart Multimodal Language Model via Chart-to-Table Pre-training and Multitask Instruction Tuning}.
\newblock In {\em Findings of the Association for Computational Linguistics: ACL 2024}.

\bibitem[Narechania et~al., 2021]{nar2021}
Narechania, A., Srinivasan, A., and Stasko, J.~T. (2021).
\newblock {NL4DV: A Toolkit for Generating Analytic Specifications for Data Visualization from Natural Language Queries}.
\newblock {\em IEEE Transactions on Visualization and Computer Graphics}, 27(2).

\bibitem[Shen et~al., 2023]{she2023}
Shen, L., Shen, E., Luo, Y., Yang, X., Hu, X., Zhang, X., Tai, Z., and Wang, J. (2023).
\newblock {Towards Natural Language Interfaces for Data Visualization: A Survey}.
\newblock {\em IEEE Transactions on Visualization and Computer Graphics}, 29(6):3121--3144.

\bibitem[Vázquez, 2024]{vaz2024}
Vázquez, P.-P. (2024).
\newblock {Are LLMs ready for Visualization?}
\newblock In {\em IEEE PacificVis 2024 Workshop - Vis Meets AI}, pages 343--352.

\bibitem[Wang et~al., 2020]{wang2020}
Wang, Y., Sun, Z., Zhang, H., Cui, W., Xu, K., Ma, X., and Zhang, D. (2020).
\newblock {DataShot: Automatic Generation of Fact Sheets from Tabular Data}.
\newblock {\em IEEE Transactions on Visualization \& Computer Graphics}, 26(1):895--905.

\bibitem[Wei et~al., 2022]{wei2022}
Wei, J., Wang, X., Schuurmans, D., Bosma, M., hsin Chi, E.~H., Xia, F., Le, Q., and Zhou, D. (2022).
\newblock {Chain of Thought Prompting Elicits Reasoning in Large Language Models}.
\newblock {\em Neural Information Processing Systems}.

\bibitem[Wu et~al., 2022]{wu2022}
Wu, A., Wang, Y., Shu, X., Moritz, D., Cui, W., Zhang, H., Zhang, D., and Qu, H. (2022).
\newblock {AI4VIS: Survey on Artificial Intelligence Approaches for Data Visualization}.
\newblock {\em IEEE Transactions on Visualization and Computer Graphics}, 28(12):5049--5070.

\bibitem[Xiao et~al., 2024]{xiao2024}
Xiao, S., Huang, S., Lin, Y., Ye, Y., and Zeng, W. (2024).
\newblock {Let the Chart Spark: Embedding Semantic Context into Chart with Text-to-Image Generative Model}.
\newblock {\em IEEE Transactions on Visualization \& Computer Graphics}, 30(1):284--294.

\bibitem[Ye et~al., 2023]{ye2023}
Ye, J., Hu, A., Xu, H., Ye, Q., Yan, M., et~al. (2023).
\newblock {UReader: Universal OCR-free Visually-situated Language Understanding with Multimodal Large Language Model}.
\newblock In {\em Findings of the Association for Computational Linguistics: EMNLP 2023}.

\end{thebibliography}

\end{document}